\documentclass[pra,twocolumn,showpacs,floatfix,superscriptaddress,aps]{revtex4}
\usepackage{amsmath}
\usepackage{makeidx}
\usepackage{amssymb}
\usepackage{graphicx}

\usepackage[usenames]{color}
\usepackage[normalem]{ulem}

\newcommand{\beq}{\begin{equation}}
\newcommand{\eeq}{\end{equation}} 
\newcommand{\beqa}{\begin{eqnarray}}
\newcommand{\eeqa}{\end{eqnarray}} 
\begin{document}

\title{A  dipolar  droplet bound in a trapped Bose-Einstein  condensate}

\author{Luis E. Young-S.\footnote{lyoung@ift.unesp.br} }
\author{ S. K. Adhikari\footnote{adhikari@ift.unesp.br; URL:
http://www.ift.unesp.br/users/adhikari}}
\affiliation{
Instituto de F\'{\i}sica Te\'orica, UNESP - Universidade Estadual Paulista, \\ 01.140-070 S\~ao Paulo, S\~ao Paulo, Brazil
} 

\begin{abstract}

We study the statics and dynamics  of a 
 dipolar Bose-Einstein condensate (BEC)
droplet bound by  inter-species contact interaction in a trapped 
non-dipolar BEC. Our findings are demonstrated in terms of 
stability  plots 
of a dipolar $^{164}$Dy droplet bound
in a trapped non-dipolar  $^{87}$Rb
BEC with a variable number of $^{164}$Dy atoms and the inter-species 
scattering length. A trapped non-dipolar BEC of a fixed number of atoms 
can only bind a dipolar droplet containing atoms less than a critical 
number for the inter-species scattering length between two critical 
values. 
%The dipolar droplet collapses for the inter-species attraction 
%above the upper limit and escapes to infinity for inter-species 
%attraction below the lower limit. 
The shape and size (statics) as well 
as the small breathing oscillation (dynamics) of the dipolar BEC droplet 
are studied using the numerical and variational solutions of a 
mean-field model. We also suggest an experimental procedure for achieving such a  $^{164}$Dy
droplet by relaxing the trap on the $^{164}$Dy
 BEC in a trapped binary $^{87}$Rb-$^{164}$Dy mixture.

\end{abstract}


\maketitle

\section{Introduction}

The experimental observation of a dipolar Bose-Einstein condensate (BEC)
 of $^{52}$Cr \cite{ExpCr,cr,saddle,crrev,52Cr},  $^{164}$Dy \cite{ExpDy,dy} and $^{168}$Er \cite{ExpEr} atoms
with large magnetic {dipole}
moments has opened new directions of research in cold atoms  in the quest of 
novel and interesting features related to the anisotropic long-range dipolar interaction. Polar molecules, 
with much larger (electric) dipole moment, are also being considered \cite{polar} for BEC 
experiments.
The atomic interaction in a dilute 
BEC of alkali-metal and other types of atoms (with negligible dipole moment)
is represented by  an S-wave contact 
(delta-function) potential. However, the non-local anisotropic long-range dipolar interaction 
acts in all partial waves and is attractive in certain directions and repulsive in others.

An untrapped three-dimensional (3D)  BEC with 
attractive interaction does not exist in nature due to collapse 
instability \cite{rmp}.  However, the collapse of an untrapped BEC 
can be avoided due to interspecies contact interaction
in a binary mixture with a trapped BEC. The shape 
of such a stable  droplet {bound 
 in a trapped BEC is controlled by the
inter-species interactions,}
%in a trapped BEC 
%by inter-species attraction 
%is controlled by the
%atomic interactions, 
whereas that of a trapped BEC is determined by the 
underlying trap. Here we consider such a dipolar droplet bound in  a trapped nondipolar BEC. 
The effect of the underlying atomic interaction, 
specially that of the anisotropic dipolar interaction, will easily 
manifest in such a  dipolar  droplet. These droplets will be termed quasi-free as they 
can easily move around inside
the larger trapped BEC responsible for their binding.   
By taking the trapped BEC to be nondipolar, one can easily study the effect of 
intra-species dipolar interaction on the quasi-free droplet in the absence of any 
inter-species dipolar interaction.

{
Dipolar BECs are immediately distinguishable from those with purely
contact interactions by their strong shape and stability dependence \cite{shape} on
trapping geometry. Here, we are proposing a new way to trap a
dipolar BEC in the form of a quasi-free 
dipolar droplet, using an attractive inter-species mean-field potential,
which introduces a unique trapping geometry that results in further
interesting shape and stability characteristics of dipolar BECs.}
We study the statics and dynamics of a quasi-free  dipolar  
droplet using numerical solution and variational approximation of
a mean-field  model \cite{mfb,mfb2}. For this purpose we consider 
a binary mixture of non-dipolar Rb87
and dipolar Dy164 where the trapped
non-dipolar $^{87}$Rb BEC could be in a cigar- or disk-shape.  %The dipole moments of the $^{164}$Dy
%atoms are considered polarized along the axial 
%$z$  direction. 
Among the available 
dipolar BECs \cite{ExpCr,ExpDy,ExpEr}, $^{164}$Dy atoms have the largest (magnetic) dipolar interaction strength
\cite{crrev,dy}.  The existence of stable 3D $^{164}$Dy droplets is illustrated by stability plots involving the number of atoms of the two species and the  inter-species scattering length. For a fixed number $N$(Rb) of $^{87}$Rb atoms the dipolar  droplet could be bound below a critical  number $N$(Dy) of  $^{164}$Dy atoms
between two limiting values of inter-species attraction. If the inter-species attraction is too small, 
{the  $^{164}$Dy atoms in the droplet cannot be bound } and, for a very large inter-species attraction, the droplet is destroyed by collapse instability. Usually, the dipolar droplet has the same shape 
as the trap acting on the non-dipolar BEC. However, for a large number of $^{164}$Dy atoms, as the inter-species 
attraction approaches the collapse instability, the dipolar droplet always is of cigar shape even if the external trap is disk shaped
and eventually the dipolar droplet  
collapses on the axial $z$ axis from the cylindrical configuration. For a small number of atoms, the dipolar droplet  collapses to the center maintaining its shape rather than on the axial $z$ axis.

 The variational approximation to the sizes and chemical potentials of the stationary droplets is compared with the numerical solution of the mean-field 
model. The numerical study of breathing oscillation of the stable dipolar droplet is found to be in reasonable 
agreement with a time-dependent variational model calculation. We also demonstrate a viable 
experimental way of creating 
a $^{164}$Dy droplet bound in a trapped $^{87}$Rb BEC, e.g., 
by slowly removing the trap on a   trapped binary $^{164}$Dy-$^{87}$Rb mixture, while the 
trapped $^{164}$Dy BEC evolves into a quasi-free droplet.

In Sec. II  the mean-field model for a   
dipolar BEC droplet bound in a trapped non-dipolar BEC is developed.
A time-dependent, analytic,  Euler-Lagrange Gaussian variational approximation 
of the model is also presented. 
The results of numerical calculation are shown in Sec. III.  
Finally, in Sec. IV we present a brief summary of our findings.

\section{Mean-field model for a quasi-free dipolar droplet}

We consider a binary BEC, where one of the species is dipolar and the 
other non-dipolar, interacting via  inter- and intra-species interactions with the 
mass, number, magnetic {/color{red} dipole} moment, and scattering length for the two species $ i=1,2,$
denoted by $m_i, N_i, 
\mu_i, a_i,$, respectively. The first species ($^{87}$Rb) 
is taken to be non-dipolar 
($\mu_1 = 0$) and trapped while the second species ($^{164}$Dy) 
is  dipolar ($\mu_2\ne 0$) and polarized along axial $z$ direction. 
The angular frequencies for the axially-symmetric trap on $^{87}$Rb
along $x$, $y$ and $z$ directions are taken as 
$\omega_x=\omega_y=\omega_1$ and 
$\omega_z=\lambda_1\omega_1$. The inter- ($V_{12}$)
and intra-species ($V_i$)
interactions 
for two atoms at positions $\bf r$ and $\bf r'$ are taken as
\begin{eqnarray}\label{intrapot} 
V_{12}({\bf R})= 
\frac{2\pi \hbar^2 a_{12}}{m_R}\delta({\bf R}),\quad
V_1({\bf R})= 
\frac{4\pi 
\hbar^2 a_1}{m_1}\delta({\bf R }),\\ \label{interpot} 
V_2({\bf R})= 
\frac{\mu_0\mu_2^2}{4\pi}\frac{1-3\cos^2 \theta}{|{\bf R}|^3}+\frac{4\pi 
\hbar^2 a_2}{m_2}\delta({\bf R }),
     \end{eqnarray}
where $\bf R = r-r',$ $\mu_0$ is the permeability of free space, 
$\theta$ is the angle made by the vector ${\bf R}$ with the polarization 
$z$ direction,  and $m_R=m_1m_2/(m_1+m_2)$ is the reduced mass of the two species of 
atoms. With these interactions, the coupled Gross-Pitaevskii (GP) 
equations for the binary dipolar BEC can be written as \cite{mfb2}
\begin{align}& \,
{\mbox i} \hbar \frac{\partial \phi_1({\bf r},t)}{\partial t}=
{\Big [}  -\frac{\hbar^2}{2m_1}\nabla^2
+ \frac{1}{2}m_1 \omega_1^2 
(\rho^2+\lambda^2_1{z}^2 )
\nonumber
\\  & 
+ \frac{4\pi \hbar^2}{m_1}{a}_1 N_1 \vert \phi_1({\bf r},t)\vert^2
+\frac{2\pi \hbar^2}{m_R} {a}_{12} N_2 \vert \phi_2({\bf r},t)|^2
{\Big ]}  \phi_1({\bf r},t),
\label{eq1}
\end{align}
\begin{align}
\label{eq2}
&{\mbox i} \hbar \frac{\partial \phi_2({\bf r},t)}{\partial t}=
{\Big [}  -\frac{\hbar^2}{2m_2}\nabla^2
\nonumber\\ &
+ \frac{4\pi \hbar^2}{m_2}{a}_2 N_2 \vert \phi_2({\bf r},t) \vert^2
+\frac{2\pi \hbar^2}{m_R} {a}_{12} N_1 \vert \phi_1({\bf r},t) \vert^2
\nonumber 
\\ & 
+ N_2 \frac{ \mu_0 \ {\mu}^2_2 }{4\pi}
\int V_{dd}({\mathbf R})\vert\phi_2({\mathbf r'},t)\vert^2 d{\mathbf r}' 
\Big] 
 \phi_2({\bf r},t),
\\&
V_{dd}({\mathbf R})= 
\frac{1-3\cos^2\theta}{{\mathbf R}^3},  \quad
  \rho^2=x^2+y^2, \quad {\mbox i}=\sqrt{-1}.  
\end{align}

To compare the dipolar and contact interactions, the intra-species 
dipolar interaction  is  expressed in terms of the length scale
$a_{dd}$, defined by $  a_{dd}=
\mu_0\mu_2^2m_2/(12\pi \hbar ^2    ).$
We express the strength of the dipolar 
interaction in Eq.  (\ref{eq2}) by this length scale
and transform Eqs. (\ref{eq1}) and (\ref{eq2}) 
into the following dimensionless form  \cite{mfb2}
\begin{align}& \,
{\mbox i} \frac{\partial \phi_1({\bf r},t)}{\partial t}=
{\Big [}  -\frac{\nabla^2}{2 }
+ \frac{1 }{2} (\rho^2+\lambda^2_1 z^2 ) 
+ g_1 \vert \phi_1 \vert^2 
 \nonumber\\ & \, %
+ g_{12} \vert \phi_2 \vert^2
{\Big ]}  \phi_1({\bf r},t),
\label{eq3}\\
& \,
{\mbox i} \frac{\partial \phi_2({\bf r},t)}{\partial t}={\Big [}  
-m_{12} \frac{\nabla^2}{2}
+ g_2 \vert \phi_2 \vert^2 
+ g_{21} \vert \phi_1 \vert^2 
\nonumber \\ & \,
+ g_{dd}
\int V_{dd}({\mathbf R})\vert\phi_2({\mathbf r'},t)
\vert^2 d{\mathbf r}'  
{\Big ]}  \phi_2({\bf r},t),
\label{eq4}
\end{align}
where
%\begin{align}&
$m_{12}={m_1}/{m_2},$
$g_1=4\pi a_1 N_1,$
$g_2= 4\pi a_2 N_2 m_{12},$
$g_{12}={2\pi m_1} a_{12} N_2/m_R,$
$g_{21}={2\pi m_1} a_{12} N_1/m_R,$
$g_{dd}= 3N_2 a_{dd}m_{12}.$
In Eqs. (\ref{eq3}) and (\ref{eq4}), length is expressed in units of 
oscillator length of the first species $l_0=\sqrt{\hbar/m_1\omega_1}$, 
energy in units of oscillator energy $\hbar\omega_1$, density 
$|\phi_i|^2$ in units of $l_0^{-3}$, and time in units of $ 
t_0=\omega_1^{-1}$.

Convenient analytic variational approximation to Eqs. (\ref{eq3}) and (\ref{eq4}) can be obtained with the following 
ansatz for the wave functions 
\cite{17,Santos01,pg}
\begin{align}
\phi_i({\bf r},t)=\frac{\pi^{-3/4}}{w_{\rho i}\sqrt{w_{z i}}}\exp\Big[-\frac{\rho^2}
{2w_{\rho i}^2}-\frac{z^2}{2w_{zi}^2}+\mathrm{i}\alpha_i\rho^2+\text{i}\beta_i z^2\Big]
\end{align}
where $ {\bf r}=\{\vec \rho,z   \}, {\vec \rho}=\{x,y\}$, $w_{\rho i}$ and $w_{z i}$ are the widths and $\alpha_i$ and 
$\beta_i$ are additional 
variational parameters. The effective Lagrangian for the binary system is 
\begin{align}
L&=\int d{\bf r} \frac{1}{2}\Big[
\sum_i \Big\{ {\text i}N_i(\phi_i\dot \phi_i^*- \phi_i^* \dot \phi_i)
+ N_ig_i|\phi_i({\bf r})|^4\Big\}
\nonumber \\
&+ N_1|\nabla \phi_1({\bf r})|^2
+m_{12} N_2|\nabla \phi_2({\bf r})|^2+
N_1(\rho^2+\lambda_1^2 z^2)\nonumber \\
&\times|\phi_1({\bf r})|^2\Big]+ N_1g_{12}\int d{\bf r}|\phi_1({\bf r})|^2   |\phi_2({\bf r})|^2
\nonumber \\
&
+ \frac{N_2}{2}g_{dd}\int \int V_{dd}({\bf R})|\phi_2({\bf r'})|^2   |\phi_2({\bf r})|^2 
d{\bf r}' d{\bf r}
, \\
%\end{align}
%\begin{align}
 \label{lag}
&= 
\sum_{i=1}^2\frac{N_i}{2}(2w_{\rho i}^2\dot \alpha_i + w_{z i}^2\dot\beta_i )+\frac{N_1}{2}
\biggr[\frac{1}{w_{\rho 1}^2}+\frac{1}{2w_{z1}^2}\nonumber \\
&+4w_{\rho 1}^2\alpha_1^2+
2w_{z1}^2\beta_1^2 \biggr]+\frac{N_2m_{12}}{2}\biggr[\frac{1}{w_{\rho 2}^2}+\frac{1}{2w_{z2}^2}
\nonumber
\end{align}
\begin{align}
 &+ 4w_{\rho 2}^2\alpha_2^2+
2w_{z2}^2\beta_2^2 \biggr]+\frac{N_1}{2}\biggr[w_{\rho 1}^2+\lambda_1^2
\frac{w_{z1}^2}{2}\biggr]
\nonumber \\&+\frac{N_1^2a_1}
{\sqrt{2\pi}w_{\rho 1}^2w_{z 1}}+\frac{N_2^2 m_{12}}
{\sqrt{2\pi}w_{\rho 2}^2w_{z 2}}[a_2-a_{dd}f(\kappa)]
\nonumber \\
&+\frac{CN_1N_2}{2
A B^{1/2}}, \quad 
\kappa=\frac{w_{\rho 2}}{w_{z2}},       \\
f(\kappa)&=\frac{1+2\kappa^2-3\kappa^2d(\kappa)}{1-\kappa^2}, \quad d(\kappa)=\frac{\text{atan}(\sqrt{\kappa^2-1})}
{\sqrt{\kappa^2-1}}, \nonumber
\end{align}
where $A=w_{\rho 1}^2+w_{\rho 2}^2$, $B=w_{z 1}^2+w_{z 2}^2$ and $C=4a_{12}m_1/(\sqrt \pi m_R)$. 
In these equations the overhead dot denotes time derivative.
%\begin{widetext}
The Euler-Lagrange variational equations for the widths for the effective Lagrangian (\ref{lag}), 
obtained in usual fashion \cite{pg}, 
are:
\begin{align}\label{eq10}
&
\ddot{w}_{\rho 1}+  {w_{\rho 1}}=
\frac{1}{w_{\rho 1}^3} +\frac{1}{\sqrt{2\pi}} \frac{2N_1 a_1}{w_{\rho 1}^3w_{z1}} +
\frac{C N_2 w_{\rho 1} }{ A^2 B^{1/2}}
, \\ \label{eq11}
& \ddot{w}_{z1} +\lambda_1^2 w_{z1}=
\frac{1}{w_{z1}^3}+ \frac{ 1}{\sqrt{2\pi}}
\frac{2N_1a_1}{w_{\rho 1}^2w_{z1}^2} 
+\frac{C N_2w_{z1}}{ A  B^{3/2}},\\ 
\label{eq12}
&
\ddot{w}_{\rho 2}=
\frac{m_{12}}{w_{\rho 2}^3} + \frac{N_2m_{12} [2a_2-
a_{dd} g(\kappa)]}{\sqrt{2\pi}w_{\rho 2}^3w_{z2}}+
\frac{C N_1 w_{\rho 2}  }{ A^2B^{1/2}}
,
 \\\label{eq13}&  \ddot{w}_{z2}  = 
\frac{m_{12}}{ w_{z2}^3} 
+\frac{2N_2m_{12} [a_2-
a_{dd}  h(\kappa)]}{\sqrt{2\pi}w_{\rho 2}^2w_{z2}^2}+\frac{CN_1  w_{z 2}}{
 A B^{3/2} },\\
&g(\kappa)=\frac{2-7\kappa^2-4\kappa^4+9\kappa^4d(\kappa)}{(1-\kappa^2)^2}, \\
&h(\kappa)=\frac{1+10\kappa^2-2\kappa^4-9\kappa^2d(\kappa)}{(1-\kappa^2)^2}. 
\end{align} 
The solution of the time-dependent 
Eqs. (\ref{eq10}) $-$ (\ref{eq13}) gives the dynamics of the variational approximation.

If $\mu_i$  is the chemical potential with which the stationary wave function $\phi_i({\bf r},t)$ 
propagates in time, e.g.  $\phi_i({\bf r},t)\sim \exp(-i\mu_it )\phi_i({\bf r})$, 
then the variational estimate for $\mu_i$ is:
\begin{align}
&\mu_1=\frac{1}{2}
\biggr[\frac{1}{w_{\rho 1}^2}+\frac{1}{2w_{z1}^2}\biggr]
+\frac{1}{2}\biggr[w_{\rho 1}^2+\lambda_1^2
\frac{w_{z1}^2}{2}\biggr]\nonumber 
\\& +\frac{2N_1a_1}
{\sqrt{2\pi}w_{\rho 1}^2w_{z 1}}+\frac{CN_2}{2
A B^{1/2}},
\\
&\mu_2=\frac{m_{12}}{2}\biggr[\frac{1}{w_{\rho 2}^2}+\frac{1}{2w_{z2}^2}
\biggr] \nonumber \\&+ \frac{2N_2 m_{12}}
{\sqrt{2\pi}w_{\rho 2}^2w_{z 2}}[a_2-a_{dd}f(\kappa)]
+\frac{CN_1}{2
A B^{1/2}}.
\end{align} 
The energy of the system is given by 
\begin{align}\label{energy}
&E=\frac{N_1}{2}
\biggr[\frac{1}{w_{\rho 1}^2}+\frac{1}{2w_{z1}^2}\biggr]
+\frac{N_1}{2}\biggr[w_{\rho 1}^2+\lambda_1^2
\frac{w_{z1}^2}{2}\biggr]\nonumber
\\& +\frac{N_2m_{12}}{2}\biggr[\frac{1}{w_{\rho 2}^2}+\frac{1}{2w_{z2}^2}
\biggr]+ \frac{N_1^2a_1}
{\sqrt{2\pi}w_{\rho 1}^2w_{z 1}}\nonumber \\&
+\frac{N_2^2 m_{12}}
{\sqrt{2\pi}w_{\rho 2}^2w_{z 2}}[a_2-a_{dd}f(\kappa)]
+\frac{CN_1N_2}{2
A B^{1/2}}.
\end{align}
The widths of the stationary state can be 
obtained from the solution of Eqs. (\ref{eq10}) $-$ (\ref{eq13}) setting the time derivatives of the widths 
equal to zero.  This procedure is equivalent to a minimization of the energy (\ref{energy}), provided the 
stationary state is stable and corresponds to a energy minimum.

\section{Numerical Results}

We solve Eqs. (\ref{eq3}) and (\ref{eq4}) by split-step 
Crank-Nicolson discretization scheme using a space step of 0.1 
and the time step 0.001 \cite{Santos01,CPC}. The contribution of the dipolar interaction is calculated in momentum 
space by Fourier transformation \cite{Santos01}. For both  species of atoms  $^{87}$Rb and $^{164}$Dy
 we take 
the intra-species
scattering length as
$a_i=110a_0$,  and the strength of dipolar interaction as $a_{dd}=131a_0$ \cite{rmp,ExpDy}.
The yet unknown inter-species scattering length $a_{12}$ is taken as a variable. The variation of  
$a_{12}$ can be achieved experimentally by the Feshbach resonance technique \cite{fesh}. We consider the trap frequency 
$\omega_1=2\pi \times 115$ Hz, so that the length scale $l_0\equiv
\sqrt{\hbar/m_1\omega_1}=1$ $\mu$m and time scale $t_0
\equiv \omega_1^{-1}= 1.38$ ms.

\begin{figure}[!t]
\begin{center}\includegraphics[width=\linewidth]{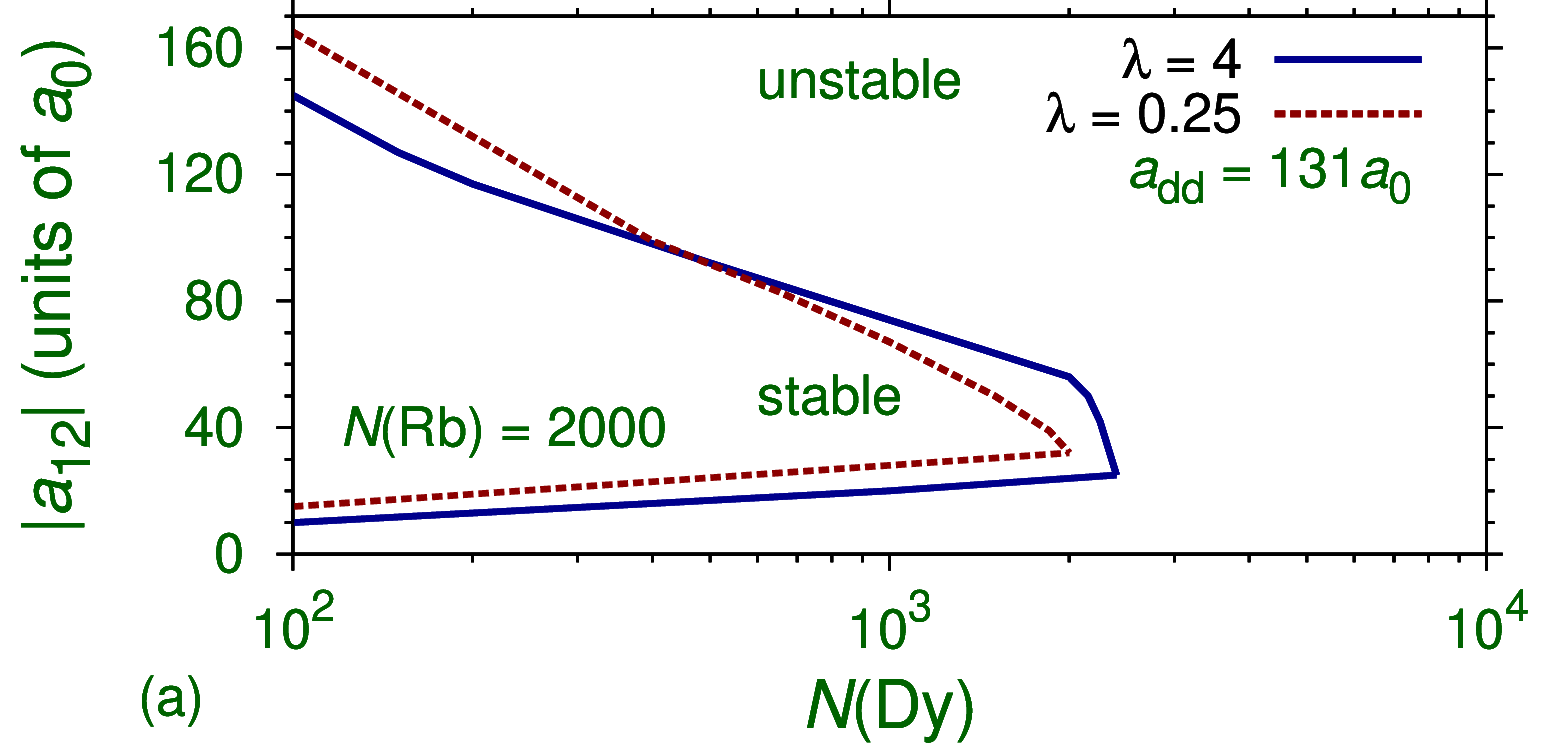}
\includegraphics[width=\linewidth]{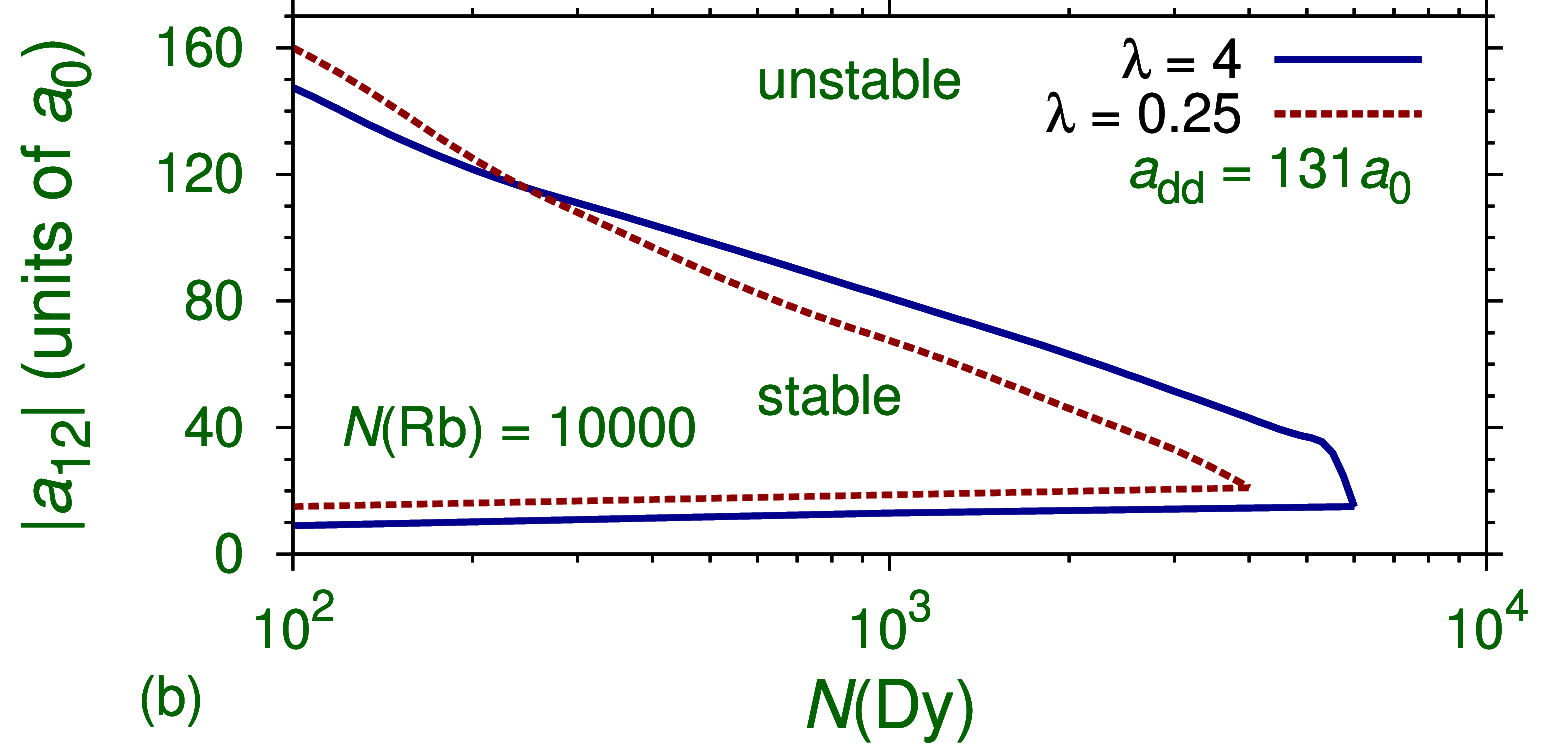}
\includegraphics[width=\linewidth]{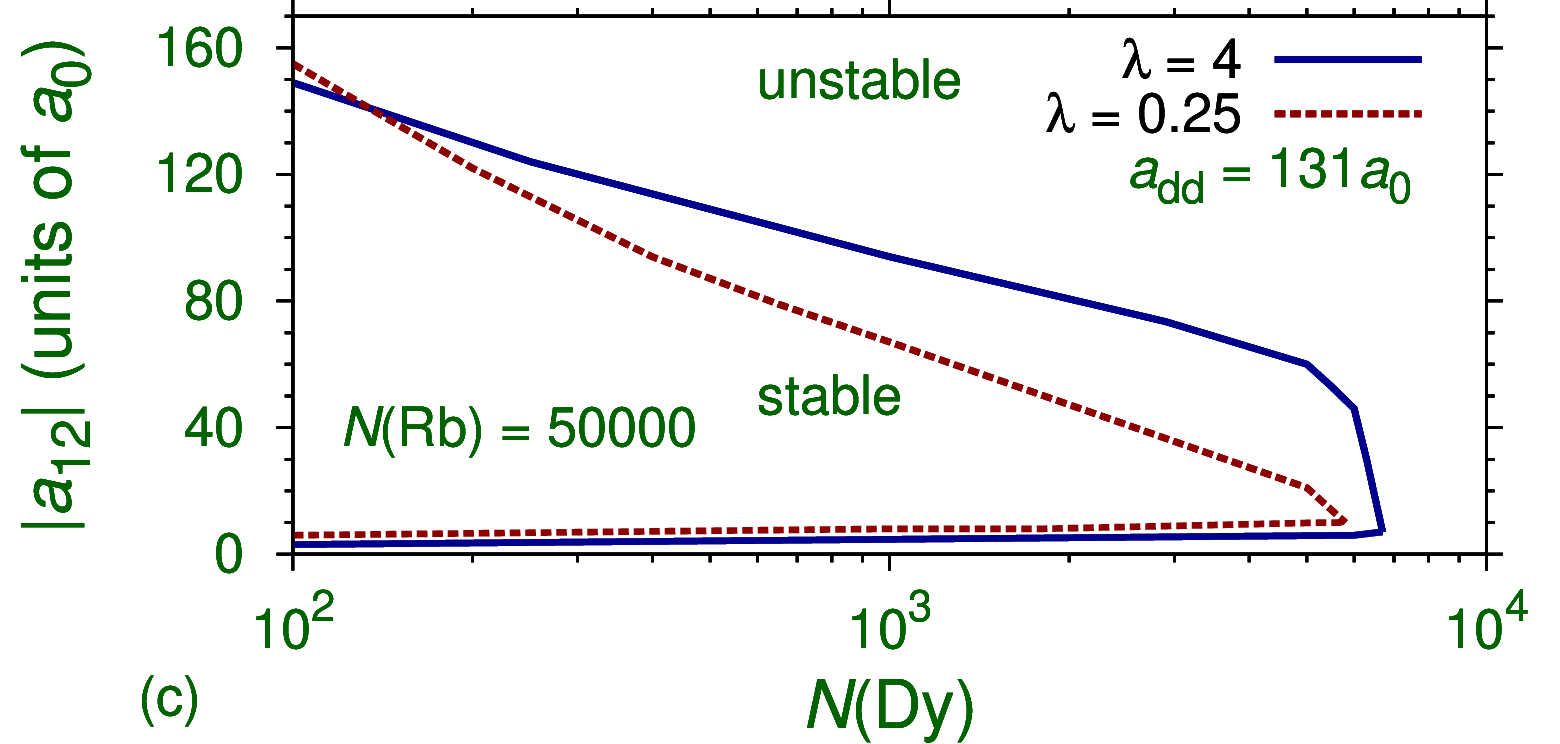}
\caption{ (Color online) $|a_{12}|-N$(Dy) stability plot showing the domain of 
stable $^{164}$Dy droplet in a binary $^{87}$Rb-$^{164}$Dy mixture
for  (a)  $N$(Rb) = 2000, (b)  $N$(Rb) = 10000, and (c)
  $N$(Rb) = 50000, and for $\lambda=0.25 $ and 4. The intra-species scattering 
lengths are taken as $a_i=110a_0$.
  }\label{fig1}
\end{center}
\end{figure}

We find that a  quasi-free  $^{164}$Dy droplet
is achievable for a moderately attractive inter-species attraction (negative $a_{12}$) and  for 
appropriate values of the number of atoms of the two species $N_1$  and $N_2$. We illustrate the 
domain of existence of a stable 
$^{164}$Dy droplet  in terms of stability plots in Figs. \ref{fig1} (a), (b) 
and (c) for 
 $N$(Rb) $=2000, 10000$ and 50000, and for $\lambda =4$ (disk-shaped trap) and 0.25 
(cigar-shaped trap). We consider cigar- and disk-shaped $^{164}$Dy
droplets in this study, where the effect of dipolar interaction is expected to be 
more prominent,  and not spherically symmetric ones where the effect of dipolar interaction is expected 
to be a minimum \cite{crrev,52Cr}.  
 In all plots of Figs. \ref{fig1}, for a fixed   $N$(Rb), 
the $^{164}$Dy droplet can be bound 
for a maximum of $N$(Dy). This maximum   of  $N$(Dy) increases with
 $N$(Rb), all other parameters remaining fixed. For a fixed 
$N$(Rb), the disk-shaped trap can accommodate a larger maximum 
number of $^{164}$Dy atoms than the  
cigar-shaped trap.
For $N$(Dy) smaller than this maximum number, 
the $^{164}$Dy 
droplet can be bound for $|a_{12}|$ between two limiting values. For $|a_{12}|$ 
above the upper limit, there is too much inter-species attraction on the 
the $^{164}$Dy droplet leading to its 
collapse.
For $|a_{12}|$ below the 
lower limit, there is not enough attraction and the $^{164}$Dy droplet {cannot be 
bound}. 
The lower limit of $|a_{12}|$ is small and tends to zero as  $N$(Rb) tends to infinity.
For  fixed values of $N$(Rb) and $N$(Dy) and for small $|a_{12}|$ near the lower limit,  
the disk-shaped configuration is favored and it can bind the $^{164}$Dy droplet  for smaller 
values of $|a_{12}|$.
However, for a fixed  $N$(Rb)  and for a large $|a_{12}|$ near the 
upper limit,  the disk-shaped configuration is favored only for a large  $N$(Dy) 
allowing a $^{164}$Dy droplet for larger $|a_{12}|$, whereas the cigar-shaped  configuration is 
favored for a small $N$(Dy) 
allowing a $^{164}$Dy droplet for larger $|a_{12}|.$

\begin{figure}[!t]

\begin{center}
\includegraphics[width=\linewidth]{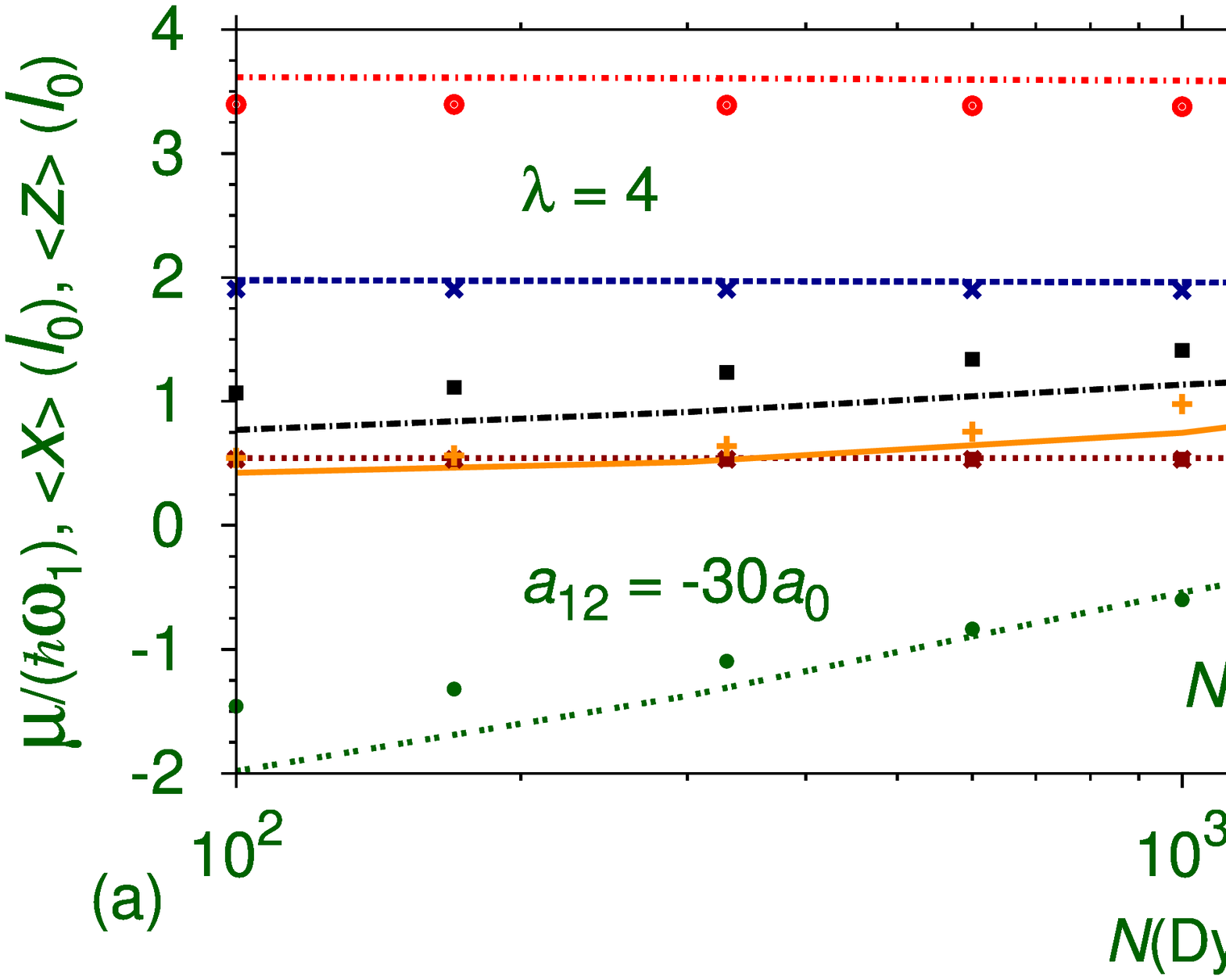}
\includegraphics[width=\linewidth]{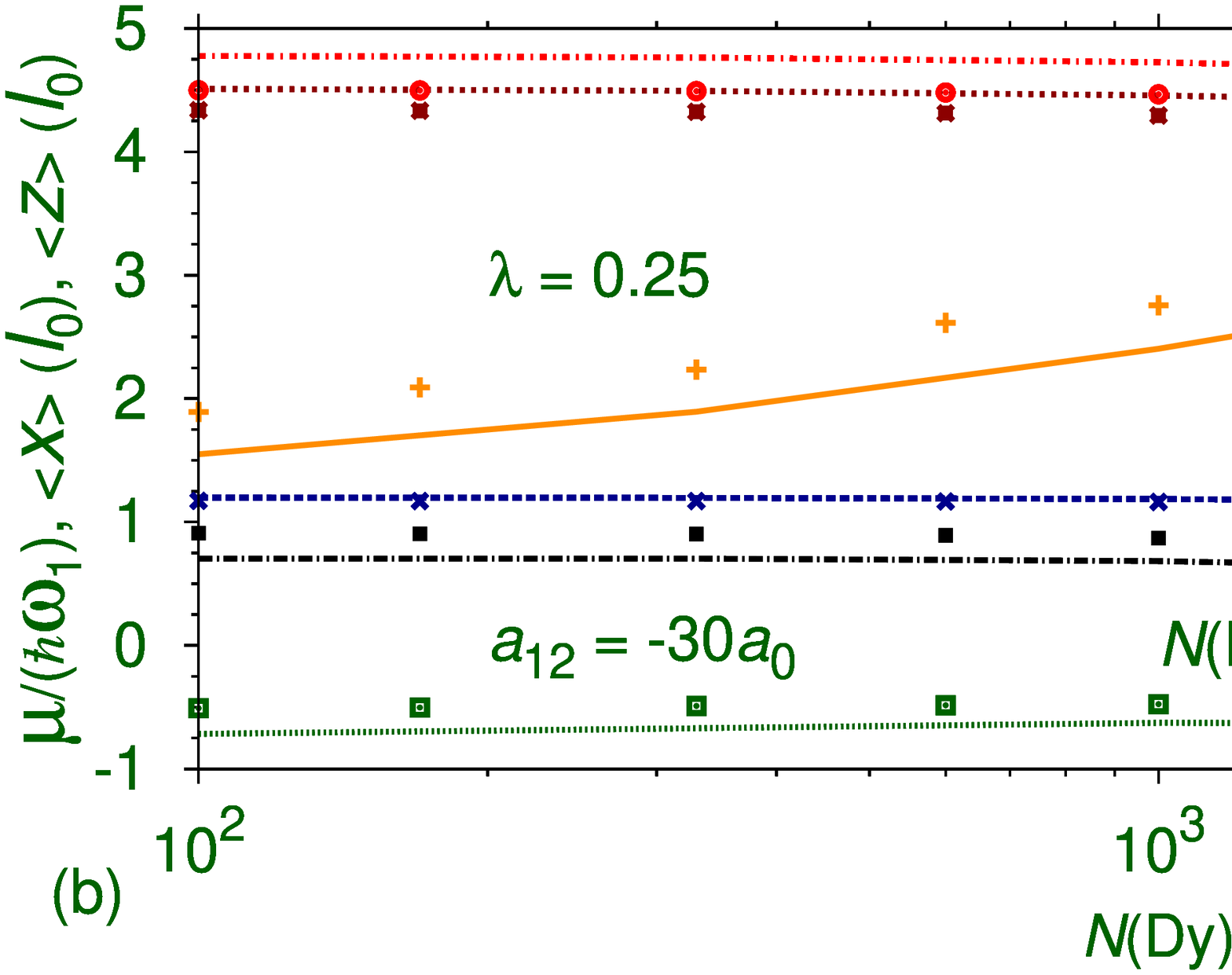}
\caption{ (Color online)  Variational (v) and numerical (n) results for chemical potential 
$\mu$ and rms sizes $\langle x \rangle,\langle z \rangle$ for the trapped $^{87}$Rb 
BEC of 10000 atoms and the bound $^{164}$Dy droplet  versus the number of $^{164}$Dy atoms for 
(a) $\lambda =4$ and (b)  $\lambda =0.25$. The inter-species scattering length is $a_{12}
=-30a_0$.  }\label{fig2}
\end{center}

\end{figure}

In Figs. \ref{fig2} we illustrate the numerical and variational results for 
chemical potentials and root-mean-square (rms) sizes $\langle x \rangle $ and 
$\langle z \rangle$ of the trapped $^{87}$Rb BEC of 10000 atoms and of the bound
 $^{164}$Dy droplet versus $N$(Dy)  for (a) $\lambda =4$ and (b) 0.25. Considering that the  $^{164}$Dy 
droplet may not have a Gaussian shape as assumed in the variational approximation, the agreement between 
the numerical and variational results is quite satisfactory.

\begin{table}
\label{I}
\caption{Root-mean-square sizes of the binary $^{87}$Rb-$^{164}$Dy system   of 50000 $^{87}$Rb atoms
from  variational approximation (\ref{eq10}) $-$ (\ref{eq13}) (var), and 
an approximate energy minimization (approx).  }
\centering
\label{table:1}
%\begin{ruledtabular}
\begin{tabular}{lrrcccccc}
\hline
&$a_{12}$ & $N$(Dy)   &$\lambda$ & $\langle x_1 \rangle$ & $\langle z_1 \rangle$   & $\langle x_2 \rangle$  & 
 $\langle z_2 \rangle$   &
   \\
\hline
approx&  -40$a_0$  & 1000 & 4 &  2.7597  & 0.7106     &1.0755  &    0.4847   \\
var& -40$a_0$ & 1000 & 4& 2.7475 &0.7091 &1.0703  & 0.4834   \\
approx& -80$a_0$ & 1000  & 0.25 &  2.2780 & 8.9525 & 0.6468 &   2.5799 \\
 var& -80$a_0$   & 1000 &0.25 &2.2549 &  8.8558 &0.6378 &  2.5525   \\
approx & -60$a_0$ &500   & 4&  2.7597 & 0.7106 & 0.8478   &  0.3768   \\
var& -60$a_0$   & 500 &4 &2.7492 & 0.7090  &  0.8443&   0.3757  \\
approx & -100$a_0$ & 100  & 0.25 &2.2780 &8.9525 & 0.6384 & 1.5803 \\
var& -100$a_0$   &100  &0.25 & 2.2752& 8.9392& 0.6373 & 1.5779   \\
\hline
\end{tabular}
%\end{ruledtabular}
\end{table}

The existence of the quasi-free $^{164}$Dy droplet can be studied qualitatively 
by a minimization of the energy (\ref{energy}).
However, the widths of the $^{87}$Rb BEC for a fixed  $N$(Rb)  do not vary much as  
  $N$(Dy)  or  $a_{12}$ is varied. Hence for a 
qualitative understanding of the 
existence of a stable $^{164}$Dy droplet for a fixed  $N$(Rb), we can make further approximation 
in the energy (\ref{energy})
and take the widths of the trapped $^{87}$Rb BEC 
to be the same as the widths of the $^{87}$Rb BEC in the absence of $^{164}$Dy atoms under otherwise
identical conditions. 
The widths of the $^{164}$Dy droplet so 
obtained for $N$(Rb) = 50000 are compared in Table I for several values of $a_{12}$, $N$(Dy), and 
$\lambda$ with the widths
obtained from exact energy minimization. We find from Table I that 
the sizes of the $^{87}$Rb for a fixed $N$(Rb) 
remains reasonably constant with respect to the variation of $N$(Dy) and $a_{12}$ and that the approximate 
energy minimization with fixed widths for $^{87}$Rb BEC leads to a good approximation to 
the widths of the $^{164}$Dy 
droplet.  In the numerical solution of the binary GP equations (\ref{eq3}) and (\ref{eq4})
we also verified that the shape and size of the
trapped   $^{87}$Rb BEC are fairly independent of the presence or absence of the $^{164}$Dy droplet. 
Hence in the study of the shape, size  and dynamics of the dipolar $^{164}$Dy
droplet bound in the trapped $^{87}$Rb
BEC, the trapped BEC will only have a 
passive role and we shall highlight only the shape, size and dynamics of the $^{164}$Dy droplet in the following.

\begin{figure}[!t]
\begin{center}
\includegraphics[width=.49\linewidth]{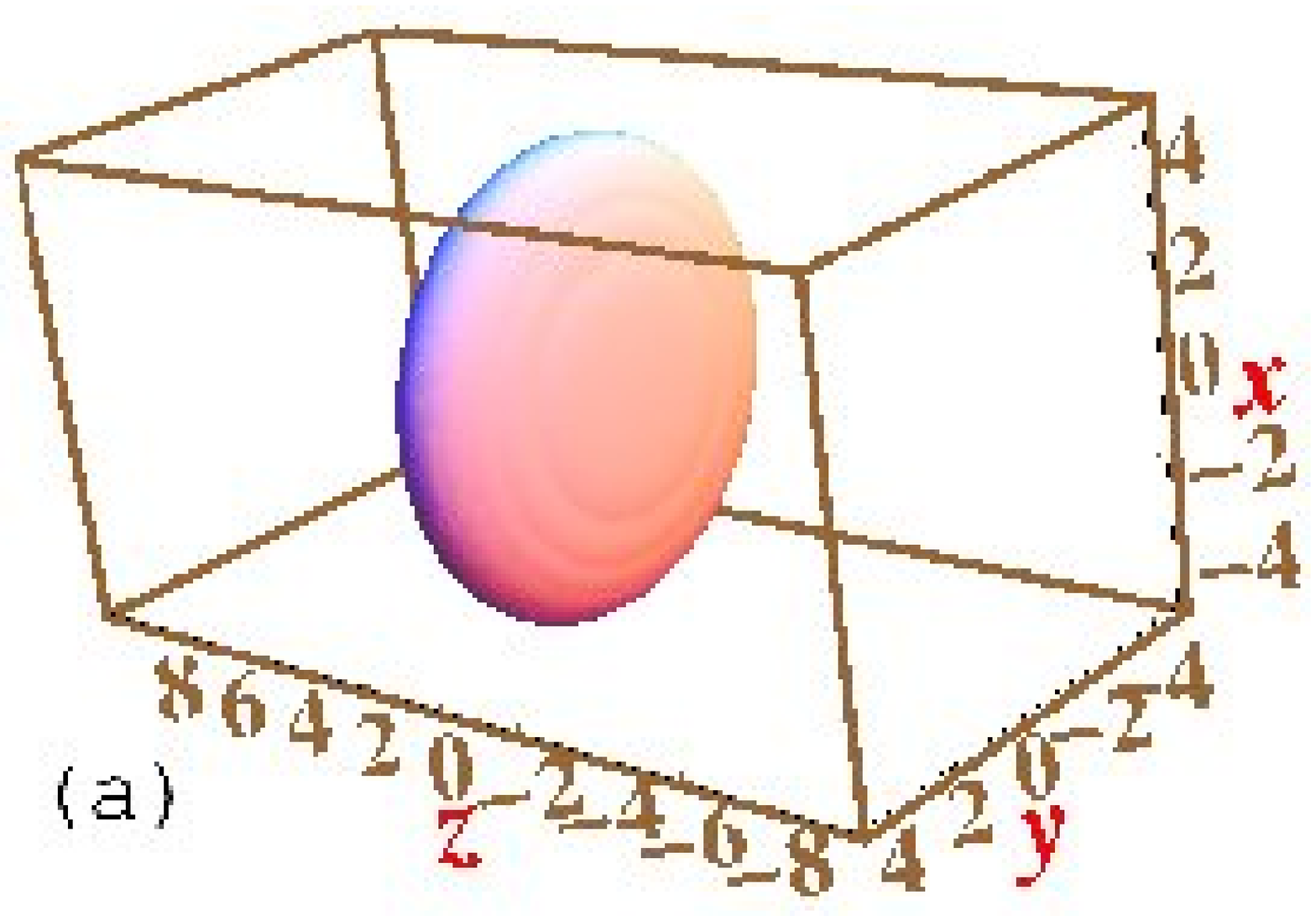}
\includegraphics[width=.49\linewidth]{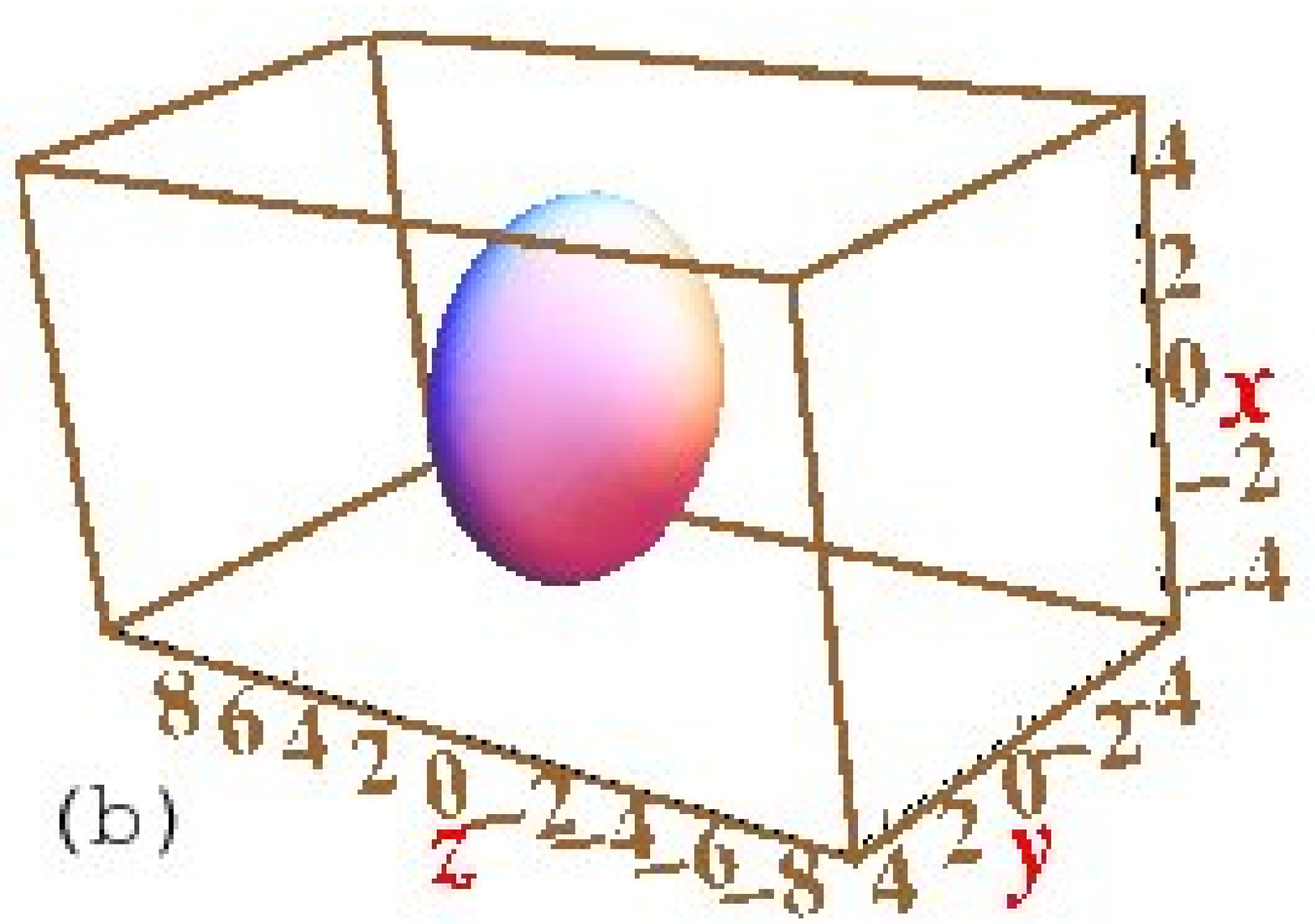}
\includegraphics[width=.49\linewidth]{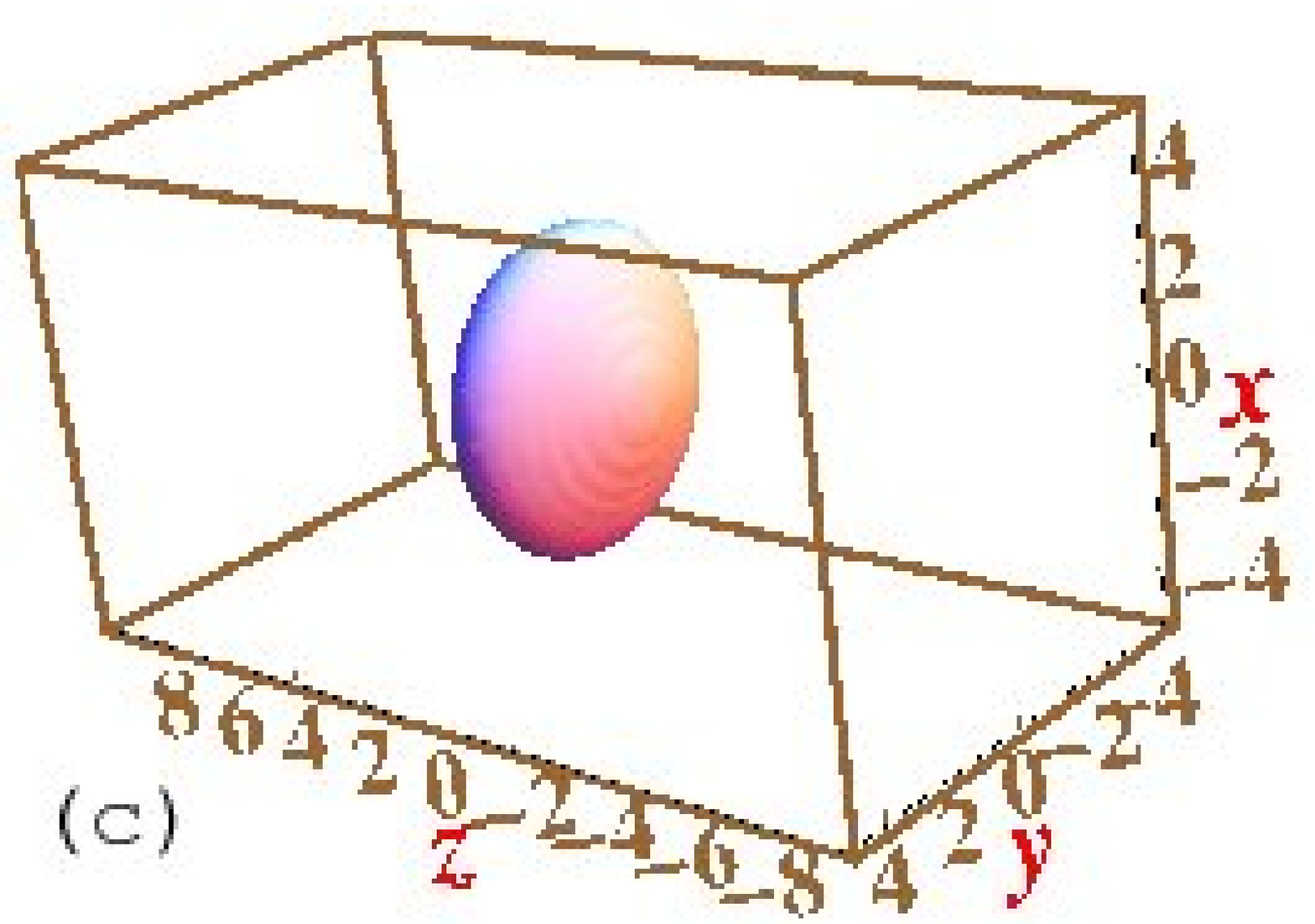}
\includegraphics[width=.49\linewidth]{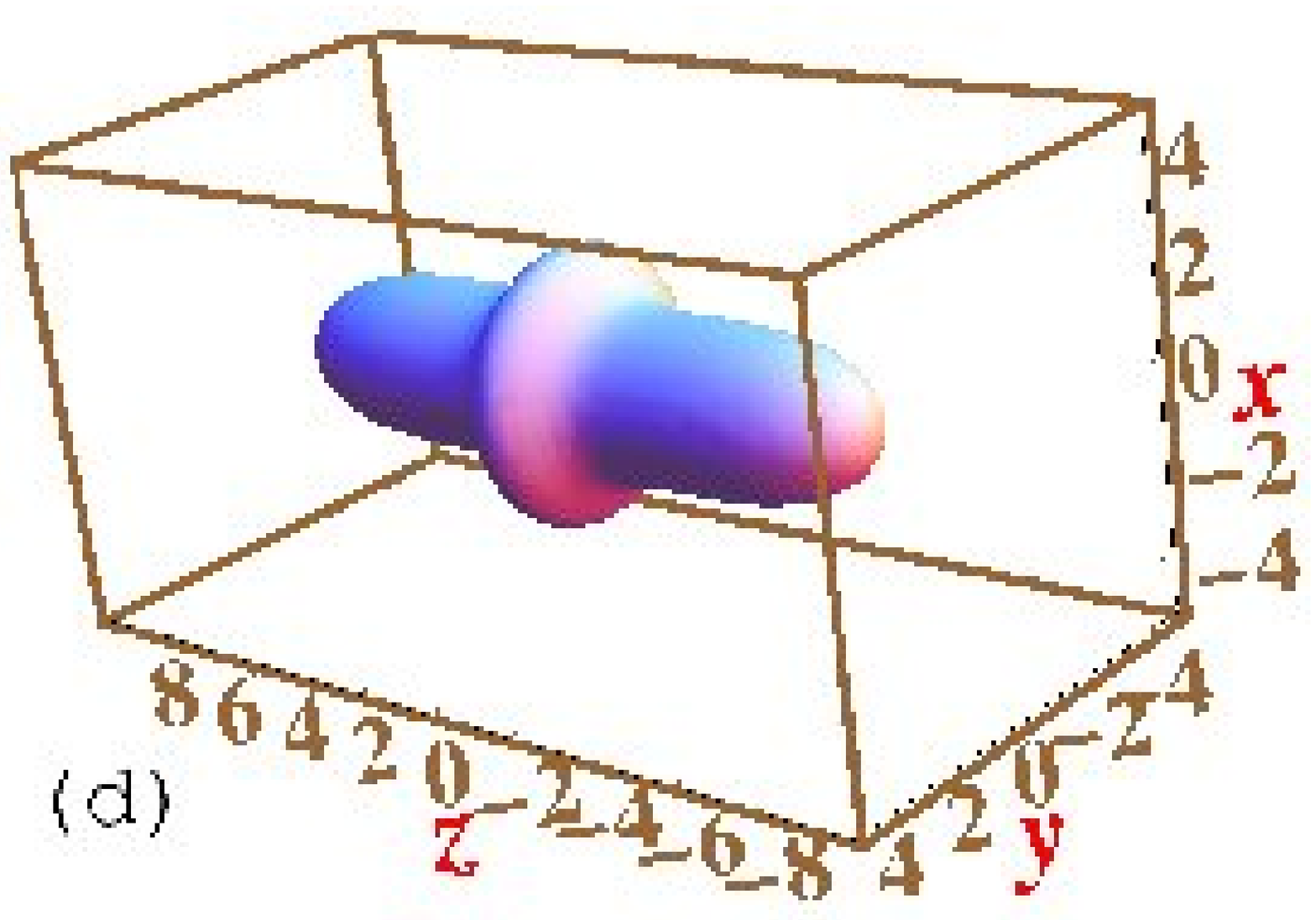}
\caption{ (Color online) (a) The  isodensity contour   of 
a disk-shaped ($\lambda=4$)  $^{87}$Rb  BEC of 10000 atoms and that 
of the  bound  $^{164}$Dy droplet of  
(b) 1000 atoms ($a_{12}=-30a_0$),
(c) 1000 atoms ($a_{12}=-80a_0$),
(d) 4000 atoms ($a_{12}=-30a_0$). The isodensity
contour of the  $^{87}$Rb  BEC is practically the same in all 
three cases.     The density on the contour in all cases is 0.001$l_0^{-3}$ and 
length is measured in units of $l_0 $ (= 1 $\mu$m).}\label{fig3}
\end{center}

\end{figure}

We consider the shape of a stable $^{164}$Dy droplet  in the stability plots of Figs. \ref{fig1} and study 
its change  as the collapse instability is approached. In Fig. \ref{fig3} (a) we show 
the isodensity contour plot  (density $|\phi_1({\bf r})|^2$)
of a
disk-shaped ($\lambda=4$)  $^{87}$Rb  BEC of 10000 atoms and 
that (density $|\phi_2({\bf r})|^2$)
of the 
bound $^{164}$Dy droplet  in    Figs. \ref{fig3} 
(b) for $N$(Dy) = 1000  ($a_{12}=-30a_0$),
(c) for  $N$(Dy) = 1000 ($a_{12}=-80a_0$), and in
(d) for $N$(Dy) = 4000  ($a_{12}=-30a_0$). The isodensity of the  $^{87}$Rb  BEC is practically
the same in all 
three cases.  Of the isodensities of the  $^{164}$Dy droplet, the one in Fig. \ref{fig3} (b) is deep  inside the 
stability region far away from collapse instability. The parameters of  Figs.  \ref{fig3} (c) and 
 (d) 
are close to the region of collapse instability. Of these two, Fig.  \ref{fig3} (c) corresponds to a medium 
number of $^{164}$Dy atoms and Fig. \ref{fig3} (d) to a  large number. 
Independent of the initial shape, a $^{164}$Dy droplet with a medium number of $^{164}$Dy atoms always collapses towards the center with shrinking size. However,  a $^{164}$Dy droplet containing a large number of atoms, independent of the associated trap symmetry, 
always   first takes  a cigar shape and then collapses on the axial $z$ direction. 
A strong dipolar interaction 
prohibits a collapse to center of a  large $^{164}$Dy droplet and favors a cigar shape. 
In Figs. \ref{fig3} the trap is disk-shaped.  The medium-sized $^{164}$Dy droplet of Fig. \ref{fig3} (c) collapses to center maintaining the disk shape
as the net dipolar interaction is smaller in this case. The  large $^{164}$Dy droplet in
Fig. \ref{fig3} (d) 
has changed its shape from the 
supporting disk-shaped trap to a cigar shape. If the attraction is further increased by increasing $|a_{12}|$, 
the $^{164}$Dy droplet of Fig.  \ref{fig3} (d) would collapse from the cigar shape on the axial $z$ axis.

\begin{figure}[!t]

\begin{center}
\includegraphics[width=.49\linewidth]{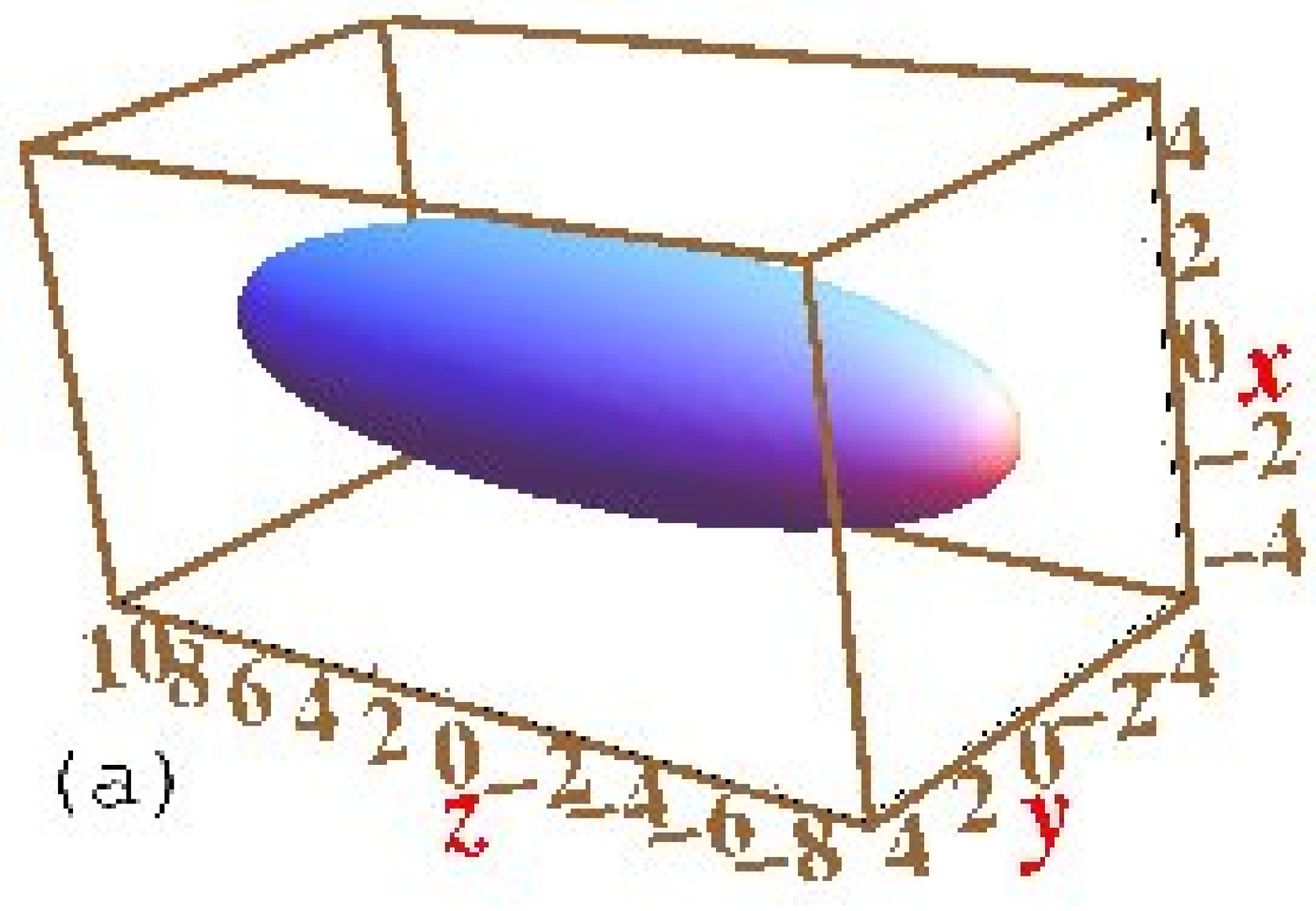}
\includegraphics[width=.49\linewidth]{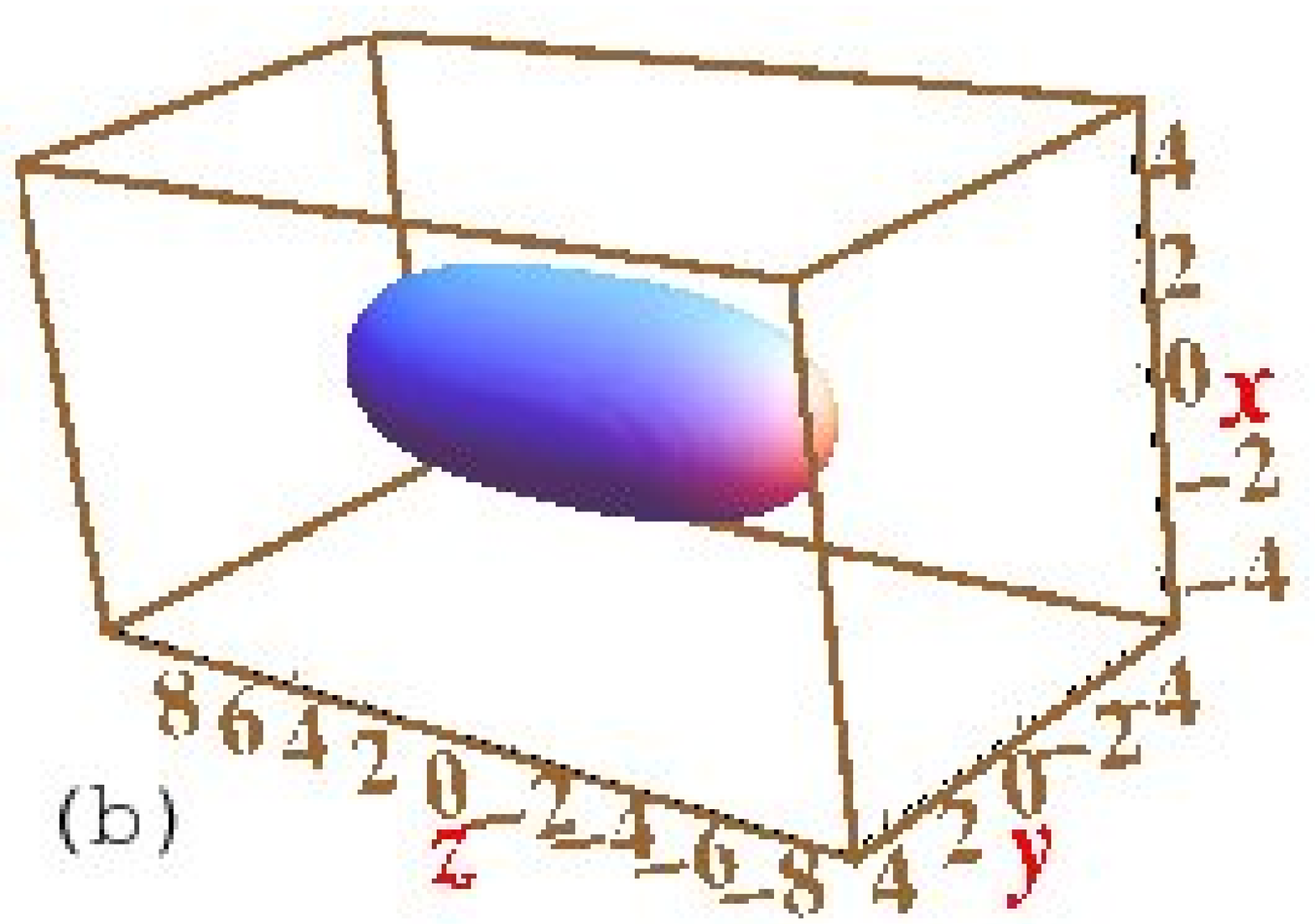}
\includegraphics[width=.49\linewidth]{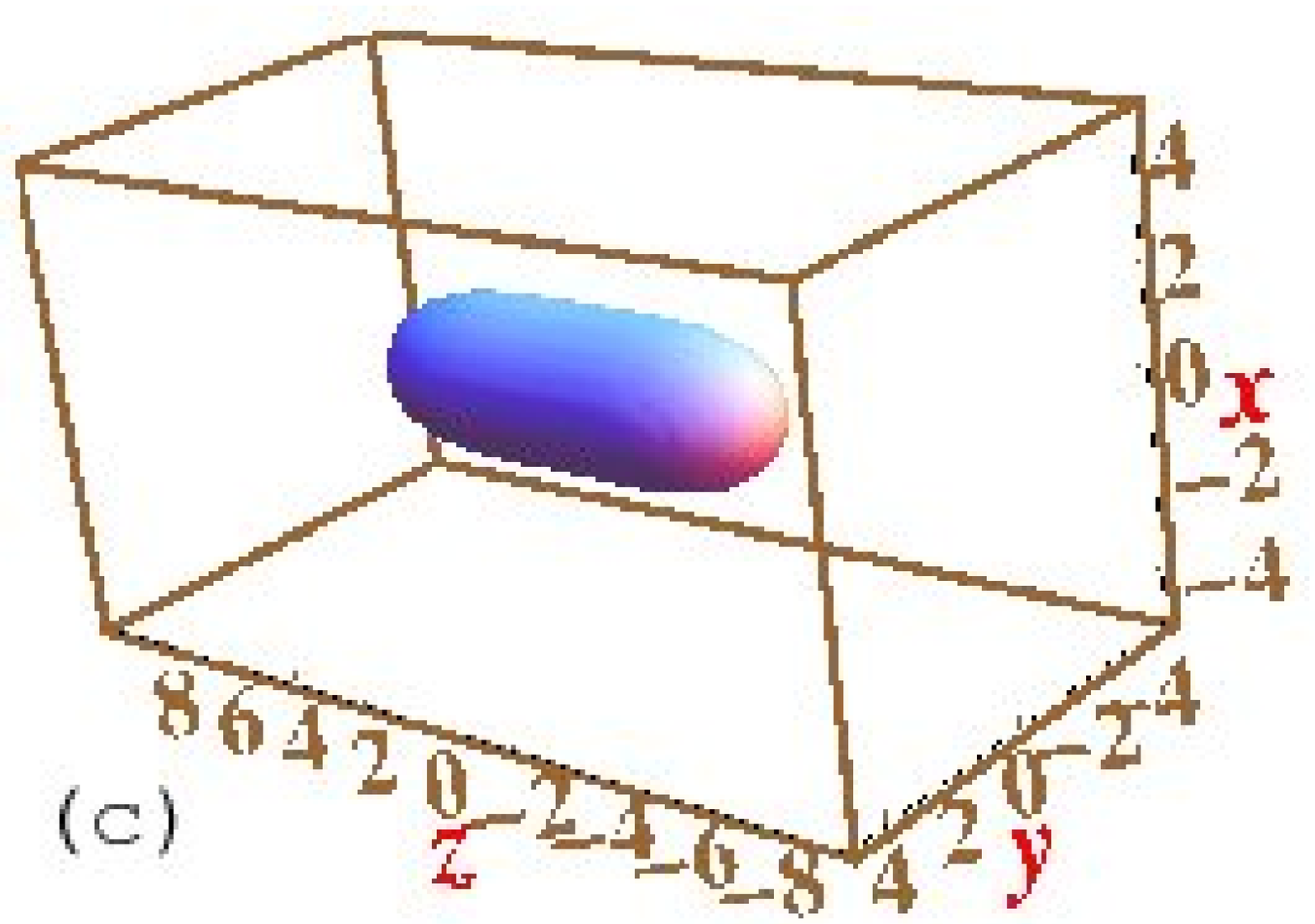}
\includegraphics[width=.49\linewidth]{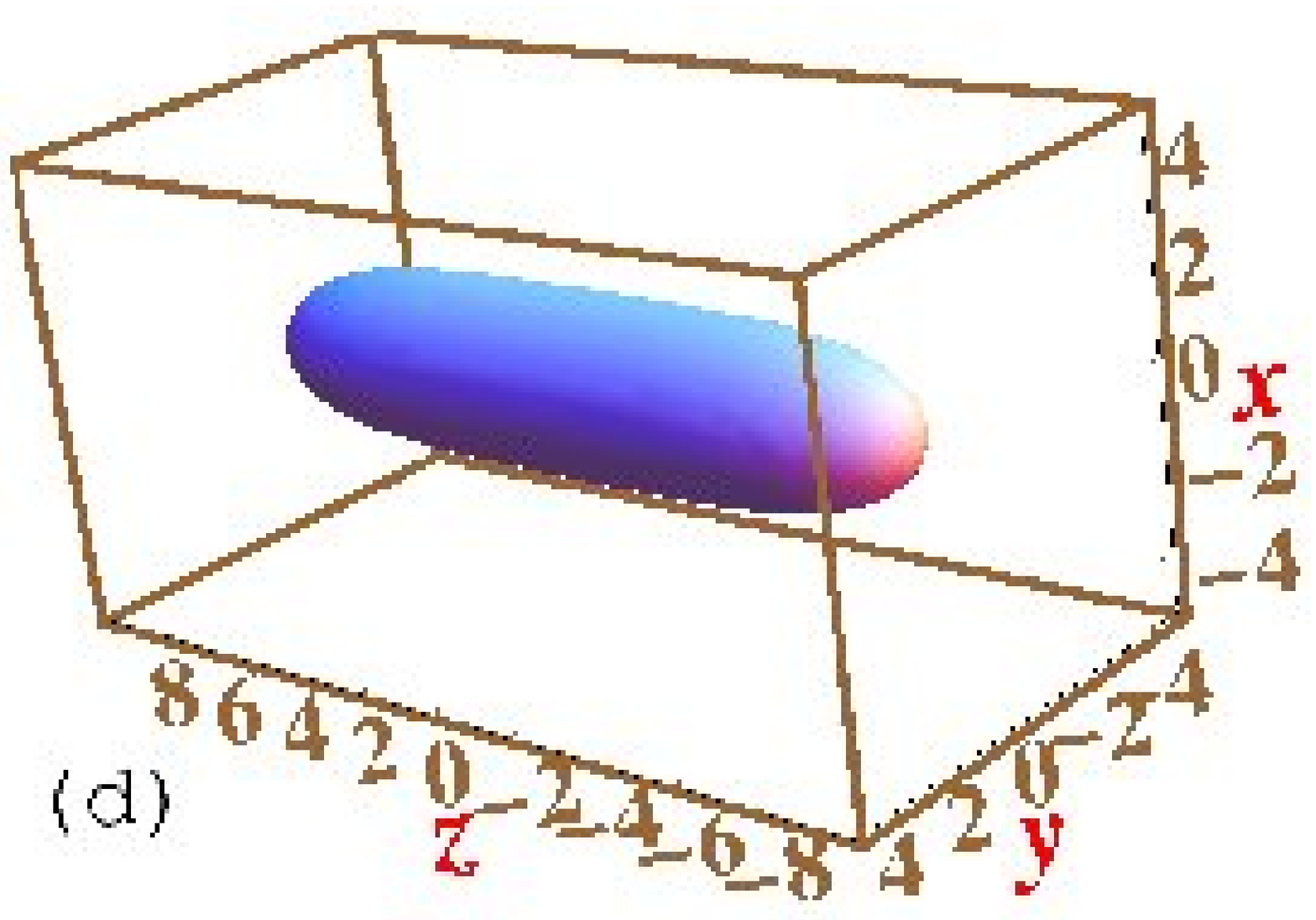}
\caption{ (Color online)(a)  The isodensity contour  of 
a cigar-shaped ($\lambda=0.25$)  $^{87}$Rb  BEC of 10000 atoms and 
that of the bound   $^{164}$Dy droplet of  
(b) 1000 atoms ($a_{12}=-30a_0$),
(c) 1000 atoms ($a_{12}=-65a_0$),
(d) 3000 atoms ($a_{12}=-30a_0$). The isodensity contour
of the  $^{87}$Rb  BEC is practically 
the same in all 
three cases.    The density on the contour in all cases is 0.001$l_0^{-3}$ and 
length is measured in units of $l_0 $ (= 1 $\mu$m).  }
\label{fig4}\end{center}

\end{figure}

\begin{figure}[!t]

\begin{center}
\includegraphics[width=\linewidth]{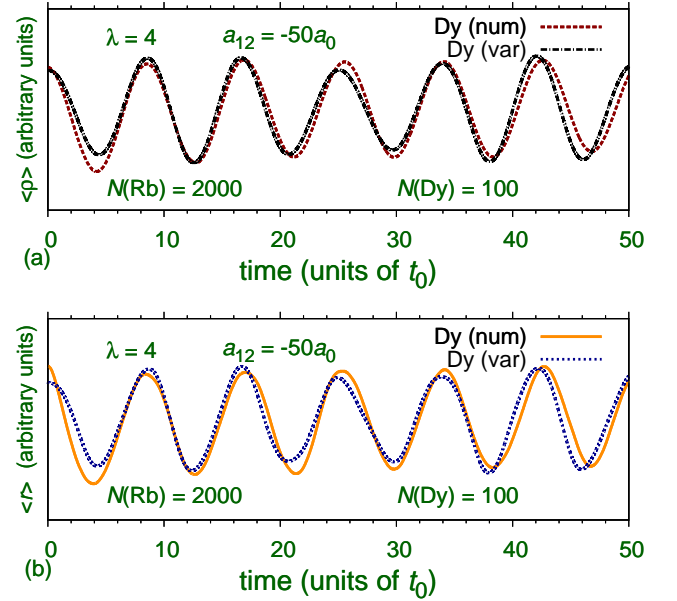}
\caption{ (Color online)  The rms sizes (a) $\langle \rho \rangle$ and (b) $ \langle r \rangle$
from numerical simulation (num) and variational approximation (var)
during breathing oscillation initiated by the sudden change $a_{12} \to 1.01\times a_{12}$ 
 versus time in units of $t_0$ (= 1.38 ms)
of a $^{164}$Dy droplet of 100 atoms bound in a trapped 
$^{87}$Rb BEC of 2000 atoms for $a_{12}=-50$ and   $\lambda =4$. }
\label{fig5}\end{center}

\end{figure}

In Fig. \ref{fig4} (a) we illustrate 
the isodensity  contour of a
cigar-shaped ($\lambda=0.25$)  $^{87}$Rb  BEC of 10000 atoms and 
the same of the  $^{164}$Dy droplet are shown in    Figs. \ref{fig4}
(b) for $N$(Dy) = 1000  ($a_{12}=-30a_0$),
(c) for  $N$(Dy) = 1000 ($a_{12}=-65a_0$), and in 
(d) for $N$(Dy) = 3000  ($a_{12}=-30a_0$). The density of the  $^{87}$Rb  BEC is virtually the same in all 
three cases.  Of the three  $^{164}$Dy droplets, the one in Fig. \ref{fig4} (b) is deep  inside the 
stability region far away from collapse instability. The parameters for Figs.  \ref{fig4} (c) and (d) 
are close to the region of collapse instability.  In all cases the $^{164}$Dy droplet maintains the cigar shape 
of the accompanying trap acting on the $^{87}$Rb BEC.  The size of the $^{164}$Dy droplet in Fig. \ref{fig4} (b) is larger 
than that  of Fig. \ref{fig4} (c) with the same number of atoms due to a strong inter-species 
attraction acting on the latter. The $^{164}$Dy droplet of Fig. \ref{fig4} (d) is  larger than those in 
Figs. \ref{fig4} (b) and (c) due to a  larger number of $^{164}$Dy atoms in it.

\begin{figure}[!t]
\begin{center}
\includegraphics[width=\linewidth]{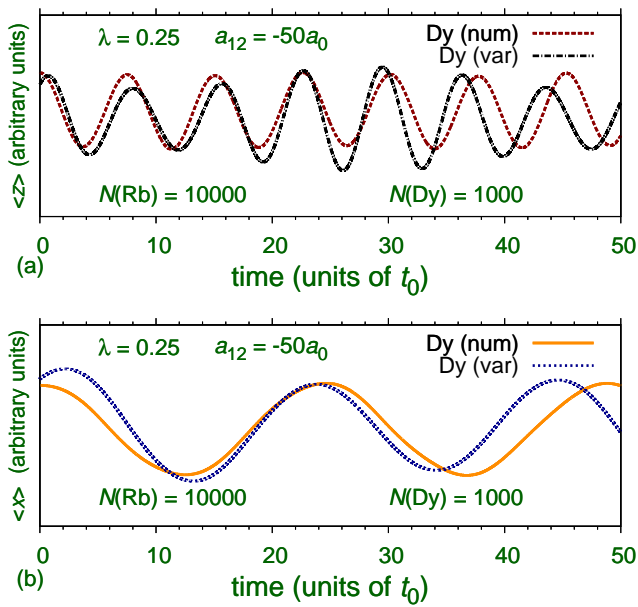}
\caption{ (Color online)  The rms sizes (a) $\langle z \rangle$ and (b) $ \langle x \rangle$
from numerical simulation (num) and variational approximation (var)
during breathing oscillation initiated by the sudden change $a_{12} \to 1.01\times a_{12}$ 
 versus time in units of $t_0$ (= 1.38 ms)
of a $^{164}$Dy droplet of 1000 atoms bound in  a trapped 
$^{87}$Rb BEC of 10000 atoms for $a_{12}=-50$ and   $\lambda =0.25$.  }
\label{fig6}\end{center}

\end{figure}

We investigate the dynamics of a bound $^{164}$Dy droplet in a trapped $^{87}$Rb BEC. 
First, to test the present scheme 
we consider a small $^{164}$Dy droplet of 100 atoms 
bound in a 
 disk-shaped ($\lambda =4$)
 $^{87}$Rb BEC of 2000 atoms  with $a_{12}=-50a_0$. 
This  
corresponds to the stable region of Fig. \ref{fig1} (a). In real-time evolution of the 
system, the scattering length $a_{12}$ is suddenly changed to $1.01\times a_{12}$ thus starting the breathing 
oscillation. In Figs. \ref{fig5} we show the resultant oscillation of the rms 
sizes (a) $\langle \rho \rangle$  and (b) $\langle r \rangle$  during time evolution as obtained from 
numerical solution of Eqs. (\ref{eq3})  and (\ref{eq4}) and variational approximation
(\ref{eq10}) $-$ (\ref{eq13}).

After obtaining a satisfactory result of dynamics with a small system,
 we venture with  a larger system of experimental interest, e. g., 
a stable $^{164}$Dy droplet of 1000 atoms 
bound in a 
 cigar-shaped ($\lambda =0.25$)
 $^{87}$Rb BEC of 10000 atoms  with $a_{12}=-50a_0$
corresponding to the stable region of Fig. \ref{fig1} (b). 
In real-time evolution of the 
system, the scattering length $a_{12}$ is again 
suddenly changed to $1.01\times a_{12}$ thus starting the breathing 
oscillation. In   Figs. \ref{fig6} we show the resultant oscillation of the rms 
sizes (a) $\langle z \rangle$  and (b) $\langle x \rangle$  during time evolution as obtained from 
a numerical solution of Eqs. (\ref{eq3})  and (\ref{eq4}) and variational approximation
(\ref{eq10}) $-$ (\ref{eq13}). 
The agreement between the numerical simulation and 
variational approximation in both cases of breathing oscillation illustrated in Figs. 
\ref{fig5} and \ref{fig6} for disk- and cigar-shapes is quite satisfactory considering the 
facts that (i) during oscillation 
the  bound $^{164}$Dy droplet might have a density distribution different 
from a Gaussian distribution assumed in the variational approximation and   
(ii) the large non-linearity 
in the $^{87}$Rb BEC will also make its density distribution non-Gaussian.  
The stable oscillation under small 
perturbation as shown in Figs. \ref{fig5} and \ref{fig6} confirms the dynamical 
stability of the $^{164}$Dy droplet 
for both disk and cigar type environments.

\begin{figure}[!t]
\begin{center}
\includegraphics[width=\linewidth]{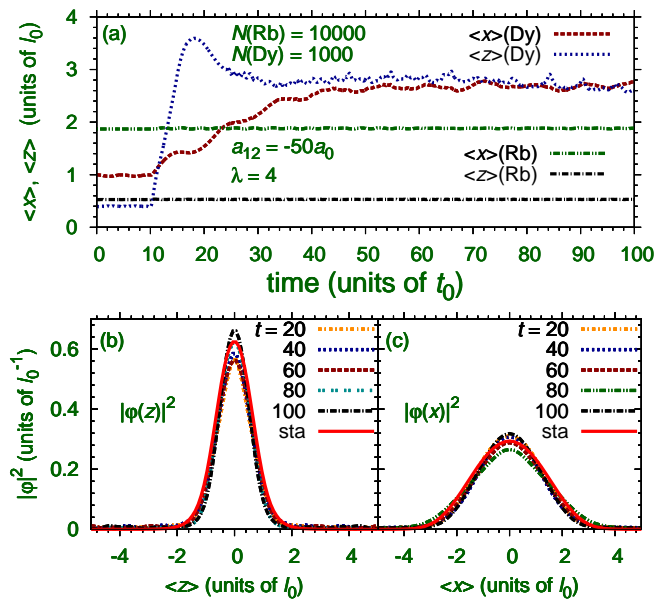}
\caption{ (Color online)  
(a) The rms sizes  $\langle x \rangle$ and  $ \langle z \rangle$ of the $^{164}$Dy and $^{87}$Rb BECs 
versus time during the passage of the trapped $^{164}$Dy BEC in the binary $^{87}$Rb-$^{164}$Dy mixture with 
$N$(Rb) = 10000, $N$(Dy) = 1000, $a_{12}=-50a_0$ and $\lambda =4$  as the confining trap on the 
$^{164}$Dy BEC is relaxed for $t > 10$ exponentially as $V_{\mathrm{trap}}\to \exp[-(t-10)]V_{\mathrm{trap}}$, so that 
this trap is practically zero for $t > 20$.
The linear 1D densities (b) $|\varphi(z)|^2$ and (c) $|\varphi(x)|^2$ at times $t=20,40,60,80,100$ 
(in units of $t_0$) during 
the expansion together with the numerically calculated density  of the stationary (sta)  $^{164}$Dy droplet.}
\label{fig7}\end{center}

\end{figure}

The present quasi-free dipolar droplet is not just of theoretical interest, but can  be realized experimentally by initially preparing a binary $^{87}$Rb-$^{164}$Dy mixture where both the components are  harmonically 
trapped. The trap on $^{164}$Dy can then be  ramped to zero exponentially during a few milli seconds. The $^{164}$Dy cloud will
 initially expand and finally form the quasi-free $^{164}$Dy droplet. To illustrate numerically 
the viability of this procedure,  we consider 
an initial  $^{87}$Rb-$^{164}$Dy binary mixture with following parameters:
$N$(Rb) = 10000, $N$(Dy) = 1000, $a_{12}=-50a_0$. We take the same axial trap with $\lambda =4 $ 
acting on both components.
Then we perform real-time propagation of the binary GP equation 
with this initial state and ramp down the $^{164}$Dy trap by $V_{\mbox{trap}}(\mbox{Dy})\to \exp[-(t-10)]V_{\mbox{trap}}$(Dy) for time $t>10$. The trap on $^{164}$Dy is practically zero for times $t\ge 20$.
The trapped $^{164}$Dy BEC first expands for $t>10$ and eventually it emerges as the quasi-free $^{164}$Dy droplet. 
The time evolution of the rms sizes of the $^{164}$Dy droplet is shown in Fig. \ref{fig7} (a). The widths of the $^{164}$Dy 
droplet first increase and 
eventually execute small 
oscillation illustrating the stable $^{164}$Dy droplet. 
%Although in Fig. \ref{fig7} (a) the $^{164}$Dy rms 
%sizes remain roughly constant during time evolution, these sizes are much larger than the sizes of the 
%corresponding stationary $^{164}$Dy droplet, which are $\langle x \rangle = 1.27$ and $\langle z \rangle = 0.65$.
 To investigate the shape of the $^{164}$Dy droplet, we plot in Figs. \ref{fig7} (b)
and (c) the one-dimensional (1D) densities $|\varphi(z)|^2\equiv \int \int dx dy |\phi(x,y,z)|^2$ and 
 $|\varphi(x)|^2\equiv \int \int dy dz |\phi(x,y,z)|^2$ at times $t=20,40,60,80,$ and 100 along with the 
corresponding numerical densities  of the stationary droplet
calculated from the solution of Eqs. (\ref{eq3}) and (\ref{eq4}). The 1D numerical densities of 
the stationary $^{164}$Dy droplet 
are in agreement with the densities obtained from dynamical simulation of the passage of the trapped $^{164}$Dy BEC to a quasi-free droplet.  In Figs. \ref{fig7} (b) and (c) one can easily identify the small oscillation 
of the evolving dynamical droplet around its stationary shape.

\section{Summary and Discussion} 

Using variational approximation and numerical solution of a set of coupled mean-field GP equations, 
we demonstrate the existence
 of a stable dipolar $^{164}$Dy droplet bound by inter-species attraction 
in a trapped non-dipolar BEC of $^{87}$Rb atoms.  The domain of stability of the 
$^{164}$Dy droplet is highlighted in stability plots of number of $^{164}$Dy atoms and inter-species scattering length
$a_{12}$ for both cigar- and 
disk-shaped traps acting on the $^{87}$Rb BEC. Results of variational approximation and numerical solution for
the statics (sizes and chemical potentials) and dynamics (breathing oscillation) of the $^{164}$Dy droplet are found to be in 
satisfactory agreement with each other. We also demonstrate numerically 
that such droplets can be obtained experimentally by 
considering a trapped binary $^{87}$Rb-$^{164}$Dy BEC and then removing the trap on $^{164}$Dy. The $^{164}$Dy BEC then expands and transforms 
into a bound dipolar droplet.

Dipolar interaction among atoms is quite different from normal short-range atomic interaction
and manifests in different ways in a trapped dipolar BEC. However, in a trapped dipolar BEC, the confining 
constraints could be too strong  and could make the effect of dipolar interaction difficult to 
observe. Special experimental set up and theoretical formulation 
might be necessary to study the effect of dipolar 
interaction. However, it will be much easier to see the effect of dipolar interaction in the 
{present binary 
mixture of non-dipolar $^{87}$Rb
and dipolar $^{164}$Dy.} Strong dipolar interaction in the axial
polarization  $z$ direction should elongate the dipolar BEC along this direction thus 
transforming it to a cigar shape.  However,  it will be difficult to observe this change in the presence 
of a harmonic trap along $z$. The effect of dipolar interaction  is clearly seen in the present 
quasi-free $^{164}$Dy droplet 
of 
Fig. \ref{fig3} (d) where due to a
strong dipolar interaction the shape of the $^{164}$Dy droplet has changed from the 
disk shape $-$  the shape of the $^{87}$Rb BEC of Fig. \ref{fig3} (a)
responsible for its binding $-$  
to the cigar shape.  The thin disk-shaped $^{87}$Rb BEC stays only near the central $z=0$ plane 
of the $^{164}$Dy droplet. Most of the $^{164}$Dy atoms lying outside the $^{87}$Rb BEC are bound due to the 
intra-species dipolar interaction. In the dipolar $^{164}$Dy droplet of Fig. \ref{fig3} (d)
the dipolar interaction is playing a more important role in its binding and shape.

{The present quasi-free dipolar droplet may also find other interesting applications in the 
BEC phenomenology.}
Among the  interesting features in a trapped dipolar BEC, one can mention its
peculiar
shape  and stability properties \cite{shape},  
and D-wave collapse \cite{collapse},  anisotropic soliton, vortex soliton \cite{soliton} and 
vortex lattice \cite{lattice}, 
anisotropic shock and sound wave 
propagation \cite{shock}, anisotropic Landau critical velocity \cite{landau},  
stable
checkerboard, stripe, and star configurations  in a two-dimensional (2D) optical
lattice as stable Mott insulator \cite{mott} as well as super-fluid
soliton \cite{15} states. It would be of great interest to find out how these features and properties 
of a trapped dipolar BEC would manifest in
a 3D quasi-free dipolar droplet. For example, a  quasi-free dipolar droplet could be used in the 
experimental study of anisotropic sound and shock-wave propagation \cite{shock}, collapse dynamics \cite{collapse}, anisotropic 
Landau critical velocity \cite{landau}, formation of vortex dipoles and vortex lattice \cite{lattice}
 etc in a different setting of confinement which will facilitate the 
observation of the effect of anisotropic dipolar interaction.

\acknowledgments
We thank FAPESP  and  CNPq (Brazil)  for partial support.

\end{document}